\documentclass[prl,twocolumn,superscriptaddress,floatfix]{revtex4}

\usepackage{graphicx}
\usepackage{bm, amsmath, amssymb}
\usepackage{color}
\usepackage{soul}
\usepackage{pdfpages}
\usepackage{amstext}
\usepackage{amsmath}
\usepackage{amsfonts}
\usepackage{amssymb}

\begin{document}

\title{Correcting correlated errors for quantum gates in multi-qubit systems
using smooth pulse control}
\author{Xiu-Hao Deng}
\email{dengxh@sustech.edu.cn}
\affiliation{Shenzhen Institute of Quantum Science and Engineering, Southern University
of Science and Technology, Shenzhen, Guangdong 518055, China}
\affiliation{Guangdong Provincial Key Laboratory of Quantum Science and Engineering
Southern University of Science and Technology, Shenzhen, Guangdong 518055,
China}
\author{Yong-Ju Hai}
\affiliation{Shenzhen Institute of Quantum Science and Engineering, Southern University
of Science and Technology, Shenzhen, Guangdong 518055, China}
\author{Jun-Ning Li}
\affiliation{Department of Physics, City University of Hong Kong, Tat Chee Avenue,
Kowloon, Hong Kong SAR, China}
\author{Yao Song}
\affiliation{Shenzhen Institute of Quantum Science and Engineering, Southern University
of Science and Technology, Shenzhen, Guangdong 518055, China}

\begin{abstract}
In multi-qubit system, correlated errors subject to unwanted interactions
with other qubits is one of the major obstacles for scaling up quantum
computers to be applicable. We present two approaches to correct such noise
and demonstrate with high fidelity and robustness. We use spectator and
intruder to discriminate the environment interacting with target qubit in
different parameter regime. Our proposed approaches combines analytical
theory and numerical optimization, and are general to obtain smooth control
pulses for various qubit systems. Both theory and numerical simulations
demonstrate to correct these errors efficiently. Gate fidelities are
generally above $0.9999$ over a large range of parameter variation for a set
of single-qubit gates and two-qubit entangling gates. Comparison with
well-known control waveform demonstrates the great advantage of our
solutions.
\end{abstract}

\maketitle

Quantum error correction provides a path to large-scale universal quantum
computers. But it is built on challenging assumptions about the
characteristics of the underlying errors in quantum logic operations to be
statistically independent and uncorrelated \cite%
{nielsen2002quantum,marques2021logical,chen2021exponential}. In realistic
environments error sources could exhibit strong temporal or spatial
correlations \cite{preskill2012sufficient}, cannot be corrected efficiently
using close-loop quantum error correction. In a multi-qubit system, due to
massive control lines and compact geometric structure, a qubit is
unavoidably to interact with other quantum systems including qubits,
couplers, neighbor control lines \cite{fried2019assessing}, etc. This
introduces control crosstalk \cite{mundada2019suppression}, leakage \cite%
{arute2019quantum}, parameter shift, decoherence and gate errors when
operating on target qubit(s). Noises become spatially-correlated, making
qubits more fragile and gate calibrations more complicated. The simultaneous
gate error rate dramatically increases compared to isolated case as qubit
system grows \cite%
{marques2021logical,chen2020demonstration,gong2019genuine,arute2019quantum}.
To better isolate the qubits and suppress the unwanted interactions, the
state-of-art techniques rely on tunable qubits and tunable couplers \cite%
{foxen2020demonstrating,cai2021perturbation,stehlik2021tunable,xu2020high,collodo2020implementation,arute2019quantum}%
. The costs of introducing the tunability are that more control lines and
more crowded spectral. Tuning over large regions of qubits' or couplers'
frequency crosses more anti-crossing points, where diadiabatic dynamics
causes additional errors \cite%
{cai2021perturbation,stehlik2021tunable,xu2020high,collodo2020implementation,arute2019quantum}%
. These obstacles limit the applications of current qubit systems and slow
down the development of scalable quantum computers.

In the NISQ era, open-loop quantum error correction (OQEC) technique is
essentially important \cite{preskill2018quantum}. As a example of OQEC
techniques, dynamical decoupling (DD) are successfully applied to further
isolate target qubits from the environment \cite%
{du2009preserving,viola1999dynamical,zeng2018general}, hence also corrects
the correlated errors due to unwanted couplings. But it is difficult to
apply DD sequences during a quantum gate operation to increase the fidelity 
\cite{li2020tunable}. Further, dynamically-corrected gate (DCG) are proposed
to suppress small dephasing errors \cite%
{khodjasteh2010arbitrarily,khodjasteh2009dynamically,zeng2018general}.
Traditional DCG relies on sequences of operation which takes the multiple
times of gate depth increases as the expense for a better single gate.
Similar to DD, use of simple pulses, such as square or $\delta-$function,
could easily cause serious waveform distortion \cite{gustavsson2013improving}%
, which brings in additional operation errors and limits the realistic
application of DD and DCG. Smooth pulse DCG has been proposed to implement
single qubit gates \cite{barnes2015robust,zeng2019geometric,zeng2018general}
to enhance the robustness of quantum gates subject to local static noise.
How to correct the errors induced during entangling two qubits is
unexplored. Also, a most drawback of conventional DCG models is the
assumption of small errors. When errors becomes significant and
non-perturbative, DCG fails.

In this paper, we study how to dynamically suppress or correct the errors in
multi-qubit system, especially the errors induced by unwanted interactions.
We analyze these errors in two different coupling regime with reference to
the control strength. For different regime, we propose different approaches
to targeted correct the errors in the relevant regime. Our approaches are
combinations of analytical theory and numerical methods, which could give
unlimited solutions of smooth control pulses for universal gate sets. We
present some examples of our solutions which are demonstrated numerically
with high fidelities. Our approach could be applied to implement robust
gates but also single qubit gate for qubit with always-on interaction with
other qubits, as well as to resolve other issues in current stage of
multi-qubit quantum computers. Our approaches provides a set of dynamical
corrected gates for qubit even with always-on interaction. In such case,
tunable couplers are not necessarily needed for scalable quantum processors
any more.

\begin{figure}[t]
\centering\includegraphics[width=1\columnwidth]{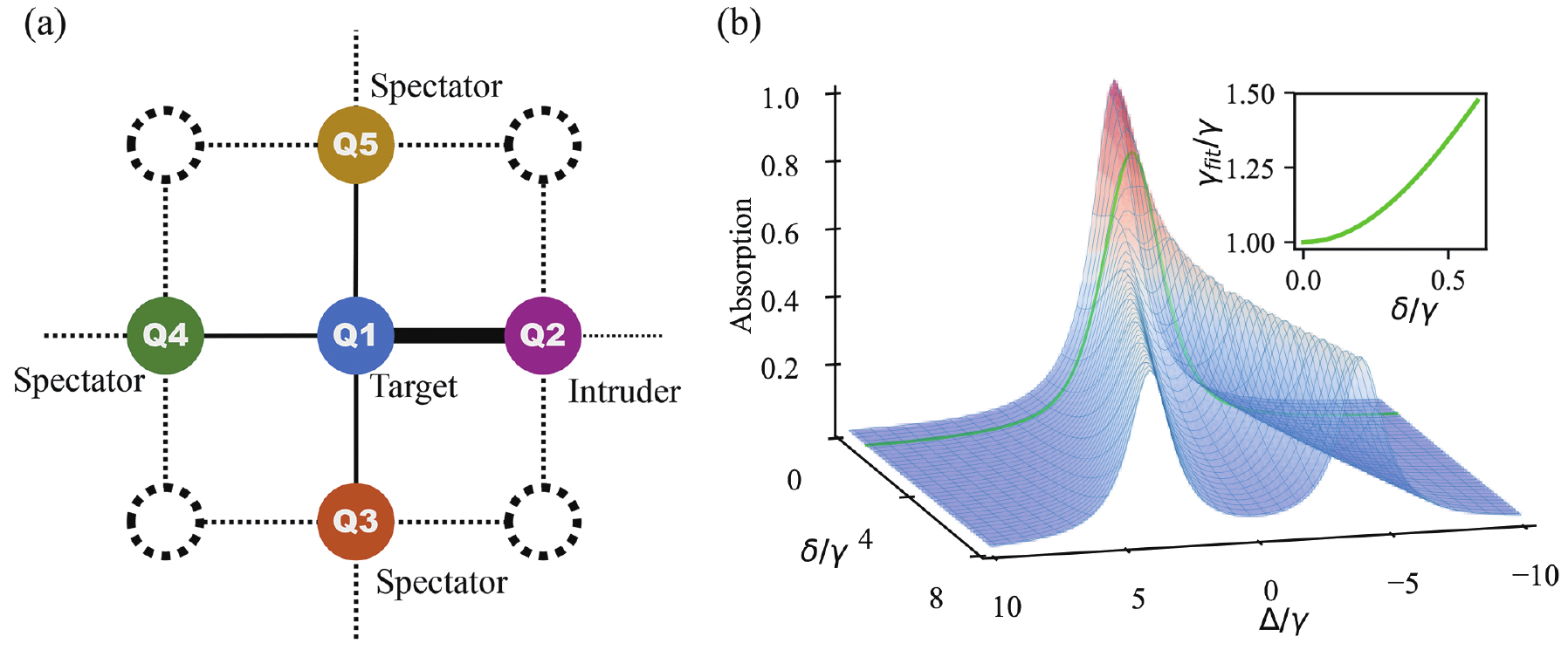}
\caption{(a) Schematic figure of multi-qubit system shows the target qubit
interacting with spectators and intruders. The thicker of the link between
qubits, the stronger residual coupling remains. (b) simulates the observed
absorption versus $\Delta$ and effective ZZ-coupling $\protect\delta$, for
non-correlated decoherence rate $\protect\gamma=1$ MHz$\times2\protect\pi$.
At $\protect\delta_{c}\approx0.7$ MHz$\times2\protect\pi$ split peaks
combine as a single peak. Hence it is the transition point between the
spectated regime intruded regime. When coupling strength is small $\protect%
\delta<\protect\delta_{c}$, the doublet's absorption overlap with each other
and combined as one broadened peak, resulting in enhanced dephasing rate.
Green line is the fitting curve of the resultant absorption profile at $%
\protect\delta=0.5$ MHz$\times2\protect\pi$. Numerical fitting of
decoherence rate $\protect\gamma^{fit}=1.341$ MHz$\times2\protect\pi$, using
a combination line shape of Lorentzian and Gaussian. The inset shows the
fitting decoherence rate enhancement versus $\protect\delta$. }
\label{Fig_MultiQB}
\end{figure}

\textit{Errors in Multi-qubit system.-}Qubits couple to each other in a
multi-qubit system, target qubit spectrum is observed to be either splitted
or broadened depending on the coupling strength, as Fig.\ref{Fig_MultiQB}(b)
shows. Its Hamiltonian in eigenbasis of a two-level subspace takes the form 
\begin{equation}
H_{t}^{diag}=-\frac{\delta}{2}\sigma_{z}  \label{Eq_Hdelta}
\end{equation}
in rotating frame with $\omega_{t}$, where $t$ stands for "target" and $%
\delta$ corresponds to leve-splitting or parameter fluctuation. $\delta$ is
generally nonvanishing in various qubit systems, including two-level system
(TLS) and multi-level system (MLS) such as superconducting transmon qubits,
as well as in various interaction regime. Analytical forms of $\delta$ in
different regime are derived in supplementary \cite{supp}.

Large interaction between qubits gives benefit to entangling operations but
harms single qubit operations. As what is used to be done in quantum dots
and superconducting qubits, detuning target qubit from the interacted qubits
could reduce the effective coupling strength but can't decouple them \cite%
{hanson2008coherent,barends2016digitized}. While effective coupling gets
much stronger than decoherence rate, target qubit's Stark splittings \cite%
{delone1999ac}\ are visible and its transition frequency in isolated-basis $%
\left\vert 0\right\rangle _{t}\longleftrightarrow\left\vert 1\right\rangle
_{t}$ splits for $\delta$ in the dressed basis (eigenbasis), corresponding
to subspace $span\{\left\vert 0\right\rangle _{i}\left\vert 0\right\rangle
_{t},\left\vert 0\right\rangle _{i}\left\vert 1\right\rangle _{t}\}$ and $%
span\{\left\vert 1\right\rangle _{i}\left\vert 0\right\rangle
_{t},\left\vert 1\right\rangle _{i}\left\vert 1\right\rangle _{t}\}$. And we
call this scenario the \textit{intruded regime} and the coupled quantum
systems are intruders, with subscript "i". Gate operations in this regime
suffer from control-rotation correlated errors. This issue is especially
serious for fix frequency qubits. The bipartite Hamiltonian is diagonalized
as $H_{t}^{diag}=diag\{\{-\frac{\omega}{2},\frac{\omega}{2}\},\{-\frac {%
\omega+\delta}{2},\frac{\omega+\delta}{2}\}\}\}$ as shown in supplement. The
second subspace could be transformed back to Eq.\ref{Eq_Hdelta}. Note that
when the target is coupled to multiple quantum systems, its spectrum split
to multiplets, with the Hamiltonian in the $n^{th}$ subspace of the target
qubit as $H_{tn}^{diag}=(-\frac{\omega}{2}+\delta_{n})\sigma_{z,t}$.

Current trending technique is to apply tunable couplers \cite%
{cai2021perturbation,stehlik2021tunable,xu2020high,collodo2020implementation,arute2019quantum}
to connect qubits with the cost of more control lines and complexity of chip
design. The tunability could be designed to realize very high on/off ratio,
which lead to successful demonstration of quantum supremacy \cite%
{arute2019quantum}. But with the realistic limit to the tunability and
imperfection of geometrical isolation, residual coupling is still a problem
for precise gate operations \cite%
{pinto2010analysis,stehlik2021tunable,arute2019quantum}[citations]. Small
couplings compared to decoherence makes level-splitting invisible but
broadens absorption peak, meaning a measured enhancement of decoherence
rate, see Fig.\ref{Fig_MultiQB}(b). The inset shows the increased
decoherence rate subject to the broadening of measured absorption. Also with
the tunable coupler, parameter fluctuations in the coupler could introduces
correlated noise to the two operated qubits during an entangling gate. We
call this regime the \textit{spectated regime}. The coupled quantum systems
are called spectators. 

Besides Stark effect, inhomogeneous driving strength and control crosstalk
errors could be induced via unwanted couplings to intruders or spectators.
The ladder operators, which the control field couples to, are modified in
the dressed basis. Take an example of a bipartite system $H_{0}=-\frac{%
\omega_{i}}{2}\sigma_{z}^{i}-\frac{\omega_{t}}{2}\sigma_{z}^{t}+g%
\sigma_{x}^{i}\sigma_{x}^{t}$ with the xy control in bare-basis $%
H_{st}^{d}=\Omega (t)(\sigma_{t}^{+}e^{i\omega_{d}t}+h.c.)$, the coupling
matrices in eigenbasis is 
\begin{equation}
\widetilde{\sigma}_{t}^{+}\newline
=\left[ 
\begin{array}{cccc}
&  &  &  \\ 
-\frac{\Delta+\epsilon}{\Upsilon_{+}} &  &  &  \\ 
\frac{2g}{\Upsilon_{+}} &  &  &  \\ 
& \frac{-\Delta+\epsilon}{\Upsilon_{-}} & \frac{2g}{\Upsilon_{-}} & 
\end{array}
\right]  \label{Eq_Dipole}
\end{equation}
where $\epsilon=\sqrt{4g^{2}+\Delta^{2}}$, $\Delta=\omega_{i}-\omega_{t} $, $%
\Upsilon_{\pm}=\sqrt{4g^{2}+(\sqrt{4g^{2}+\Delta^{2}}\pm\Delta)^{2}} $. $%
\Omega(t)$ drives transitions $\widetilde{\left\vert 00\right\rangle }%
\leftrightarrow\widetilde{\left\vert 01\right\rangle }$ and $\widetilde{%
\left\vert 10\right\rangle }\leftrightarrow\widetilde{\left\vert
11\right\rangle }$ with inhomogeneous dipole moments, as well as crosstalk
on transitions $\widetilde{\left\vert 00\right\rangle }\leftrightarrow 
\widetilde{\left\vert 10\right\rangle }$ and $\widetilde{\left\vert
01\right\rangle }\leftrightarrow\widetilde{\left\vert 11\right\rangle }$.

As well known that in solid state qubit system, low frequency noise is one
of the major intrinsic error sources \cite%
{yoshihara2006decoherence,bialczak20071,paladino20141}. Usually gate
operations are much shorter than qubit's life time by several orders of
magnitude, hence the low frequency nature of such noises behaves as a static
error for quantum gates. The low frequency for a single qubit takes the same
form as in Eq.\ref{Eq_Hdelta}. While for a two-qubit operation such as $%
XX+YY $ interaction, the noise $\delta$ on target qubit 2 appears in the
total Hamiltonian as this form

\begin{equation}
H^{total}=\left[ 
\begin{array}{cccc}
-\delta/2 &  &  &  \\ 
& \delta/2 & g(t) &  \\ 
& g(t) & -\delta/2 &  \\ 
&  &  & \delta/2%
\end{array}
\right]
\end{equation}
The entangling dynamics is only affected in the two-level subspace $span\{%
\widetilde{\left\vert 01\right\rangle },\widetilde{\left\vert
10\right\rangle }\}$ \cite{JiaweiQiu2021dynamical}, hence the effective
Hamiltonian again takes the same form as Eq.\ref{Eq_Hdelta} excluding the
control field $\Omega(t)=g(t)$.

\begin{figure}[t]
\centering\includegraphics[width=1\columnwidth]{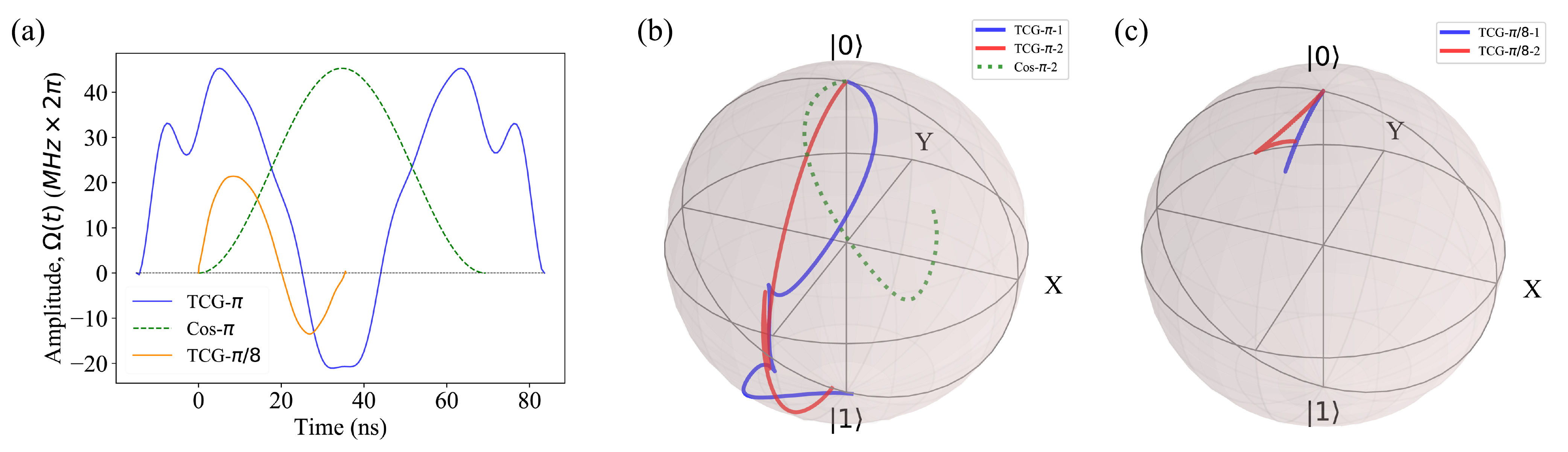}
\caption{(a) shows the waveform to implement $\protect\pi$ and $\protect\pi%
/8 $ gates at the present of an intruder, respectively in purple and orange.
As an comparison, a widely used cosine waveform for $X$ gate is shown as
dashed line. (b) shows the trajectories of $\protect\widetilde{\left\vert
0_{i}0_{t}\right\rangle }\rightarrow\protect\widetilde{\left\vert
0_{i}1_{t}\right\rangle }$ and $\protect\widetilde{\left\vert
1_{i}0_{t}\right\rangle }\rightarrow \protect\widetilde{\left\vert
1_{i}1_{t}\right\rangle }$ operated by the TCG for a $\protect\pi$
X-rotation with fidelity $0.999$, for comparison, added with a trajectory $%
\protect\widetilde{\left\vert 1_{i}0_{t}\right\rangle }\rightarrow\protect%
\widetilde{\left\vert 1_{i}1_{t}\right\rangle }$ operated by a widely used
cosine pulse, which gives a fidelity $0.589$. (c) shows the trajectories of $%
\protect\widetilde{\left\vert 0_{i}0_{t}\right\rangle }\rightarrow\protect%
\widetilde{\left\vert 0_{i}1_{t}\right\rangle }$ and $\protect\widetilde{%
\left\vert 1_{i}0_{t}\right\rangle }\rightarrow \protect\widetilde{%
\left\vert 1_{i}1_{t}\right\rangle }$ operated by the TCG for a $\protect\pi%
/8$ X-rotation with fidelity $0.9999$. These simulations use superconducting
transmon qubits. }
\label{Fig_TCG}
\end{figure}

\textit{Targeted-correction gates in intruded regime.-}In this regime, the
level splitting $\delta $ is large, seen as splitted spectrum as Fig.\ref%
{Fig_MultiQB} (b). The Hamiltonian in the $4$-by-$4$ computational subspace $%
span\{\left\vert \text{intruder}\right\rangle \left\vert \text{target}%
\right\rangle \}$ is diagonalized as $H^{diag}=Diag\{-\frac{\omega }{2}$, $%
\frac{\omega }{2}$, $-\frac{\omega +\Delta }{2}$, $\frac{\omega +\Delta }{2}%
\}$. Control-rotational error between the intruder and the target qubit is
significant if using simple pulses, see the green dashed line in Fig.\ref%
{Fig_TCG} (b). Besides, a quantum gate targeted-correcting such errors needs
to consider inhomogeneous dipole moments as shown in Eq.\ref{Eq_Dipole}. As
pointed out above, implementing two-qubit entangling gates becomes
relatively trivial in this regime compared to single qubit gates. A large ZZ
interaction directly generates a CZ gate while large XX interaction
implements an iSWAP gate. Here we present our targeted-correction gate (TCG)
approach for given $\delta $ and dipole moment inhomogeneity to implement
single qubit operations. First we decompose the $SU(4)$ dynamics to $%
SU(2)\otimes SU(2)$ dynamics. A control field $\tilde{H}_{st}^{d}=(\Omega
^{\ast }(t)\widetilde{\sigma }_{t}^{+}e^{i\omega _{d}t}+h.c.)$ is added to
the target qubit. Drive amplitude $\Omega (t)$ could be complex for IO
control [citations] meaning driving both X and Y directions. Transformed to
interaction picture with $\tilde{H}_{0}$, the unitary operator 
\begin{equation}
U^{I}(t)=\left[ 
\begin{array}{cc}
U_{1} & B \\ 
-B^{\dagger } & U_{2}%
\end{array}%
\right]   \label{Eq_blockU}
\end{equation}%
Because $B$ includes fast oscillating phases $e^{i2\omega t}$ by taking $%
\omega _{d}=\omega $, it could be demonstrated that $B=0$ under the
constraints $\delta \ll \omega $ \cite{supp}. So the evolution operator is
block diagonalized $U^{int}(t)=Diag\{U_{1},U_{2}\}$. Since the first
subspace $span\{\widetilde{\left\vert 00\right\rangle },\widetilde{%
\left\vert 01\right\rangle }\}$ is driven resonantly $\omega _{d}=\omega $, $%
U_{1}(t)=\exp \{-i\frac{\Delta +\epsilon }{\Upsilon _{+}}%
\int_{t_{0}}^{t_{0}+\tau _{g}}dt\Omega (t)\sigma _{x}\}$. For the second
subspace $span\{\widetilde{\left\vert 10\right\rangle },\widetilde{%
\left\vert 11\right\rangle }\}$, $U_{2}(t)=\mathcal{T}\exp \{-i\frac{2g}{%
\Upsilon _{-}}\int_{t_{0}}^{t_{0}+T}dt[\Omega \cos (t\delta )\sigma
_{x}-\Omega \sin (t\delta )\sigma _{y}]\}$. Generally, $U_{1}(t)\neq U_{2}(t)
$, resulting in different trajectories depending on intruder's state, see
Fig.\ref{Fig_TCG} (b) and (c). Setting $U_{1}(T)=U_{ideal}$ to get the area
constraint $\frac{\Delta +\epsilon }{\Upsilon _{+}}\int_{t_{0}}^{t}\Omega
(t)dt=\theta _{ideal}/2$, where $\theta _{ideal}$ is the ideal rotational
angle and $T$ is the ending operation time. Setting $U_{2}(T)=U_{1}(T)$ to
correct the control-rotational error, we get additional constraints $\frac{2g%
}{\Upsilon _{-}}\int_{t_{0}}^{t}\Omega (t)\cos (\delta t)dt=\theta _{ideal}/2
$ and $\int_{t_{0}}^{t}\Omega (t)\sin (\delta t)dt=0$. We then apply smooth
pulse ansatz%
\begin{equation}
\Omega (t)=\sum C_{n}\cos (2\pi \frac{nt}{T}+\phi _{n})  \label{Eq_TCGansatz}
\end{equation}%
to numerically search for appropriate waveforms for TCG of $\theta _{ideal}$
rotations. Results are shown in Fig.\ref{Fig_TCG}. On the Bloch spheres, The
quantum trajectories $\left\vert 0\right\rangle _{t}\rightarrow \left\vert
1\right\rangle _{t}$ and $\left\vert 0\right\rangle _{t}\rightarrow \cos 
\frac{\pi }{16}\left\vert 0\right\rangle _{t}+\sin \frac{\pi }{16}\left\vert
1\right\rangle _{t}$ in different subspaces $span\{\widetilde{\left\vert
00\right\rangle },\widetilde{\left\vert 01\right\rangle }\}$ and $span\{%
\widetilde{\left\vert 10\right\rangle },\widetilde{\left\vert
11\right\rangle }\}$ demonstrated the idea of targeted-correction of $\pi $
and $\pi /8$ XY gates \cite{nielsen2002quantum}. A further numerical
optimization using GRAB method \cite{Yao2021GRAB}, a GRAPE \cite%
{khaneja2005optimal} assisted analytical optimization approach, could obtain
fidelity $>0.9999$. The TCG pulse to implement the X gate shown in Fig.\ref%
{Fig_TCG}(a) with the analytical formula Eq.\ref{Eq_TCGansatz}. For $\pi $
gate, $C_{n}\in \{19.6$, $17.5$, $-22.5$, $-0.7$, $1.9$, $0.5$, $2.9$, $-4.6$%
, $-2.5$, $-2.8\}$ MHz$\times 2\pi $, and $\phi _{n}\in \{0.941$, $1.883$, $%
-0.3701$, $0.628$, $1.580$, $8.774$, $0.3506$, $1.346$, $2.128\}$, $T=98.6ns$%
. The fidelity is $0.998$. While for $\pi /8$ gate, $C_{n}\in \{4.8$, $-17.7$%
, $-0.6$, $-1.1$, $-0.5$, $-0.4$, $-0.3$, $-0.2$, $-0.1$, $-0.1\}$ MHz$%
\times 2\pi $, and $\phi _{n}\in \{1.481$, $8.663$, $0.2454$, $0.1687$, $%
0.1230$, $0.1042$, $0.0747$, $-0.0134$, $0.2745\}$, $T=35$ ns. The fidelity
is $0.999$. While allowing more harmonic components, both optimized
fidelities are demonstrated to be above $0.99999$. The numerical simulations
here are demonstrated on superconducting transmon qubits connected via a
coupler $H=\sum_{j}(\omega _{j}a_{j}^{\dag }a_{j}-\frac{\alpha _{j}}{2}%
a_{j}^{\dag }a_{j}^{\dag }a_{j}a_{j})+g_{cj}(a_{c}^{\dag
}+a_{c})(a_{j}+a_{j}^{\dag })$, all with three levels. The parameters are $%
\omega _{1,2,c}=\{5.7735$, $5.4735$, $6.99\}$GHz$\times 2\pi $, $\alpha
_{1,2}=249$ MHz$\times 2\pi $, $g_{c1,c2}=151$ MHz$\times 2\pi $. The first
subspace $span\{\widetilde{\left\vert 00\right\rangle },\widetilde{%
\left\vert 01\right\rangle }\}$ is driven resonantly.

\textit{Errors robust gates in the spectated regime.-}When the errors are
small compared to decoherence rate $\gamma$. $\delta$ is close to the
background fluctuations, hence the $\delta$ spreads over a range of $%
[-\Delta,\Delta]$. Instead of correcting a targeted value of $\delta$, now
the quantum gates should be robust to this a range of errors. This requires
a high fidelity plateau at $[-\Delta,\Delta]$ around $\delta=0$. We take the
perturbative approach of dynamically-corrected gate \cite%
{zeng2019geometric,barnes2015robust,zeng2019geometric,zeng2018general} to
find error-robust gates (ERG) in this regime \cite{YJHai2021construct}.
Consider the noisy Hamiltonian with control fields

\begin{equation}
H=\left[ 
\begin{array}{cc}
-\delta /2 & \Omega (t)e^{i\alpha (t)} \\ 
\Omega (t)e^{-i\alpha (t)} & \delta /2%
\end{array}%
\right]
\end{equation}%
The generated evolution operator $U=U_{0}U_{e}$, where $U_{0}=e^{-i%
\int_{0}^{t}H_{0}(\tau )d\tau }$ is the ideal evolution operator generated
by $H_{0}=\Omega \cos \theta \sigma _{x}-\Omega \sin \theta \sigma _{y}-%
\frac{\epsilon }{2}\sigma _{z}$. And $U_{e}=U_{0}^{\dagger
}U=I-i\int_{0}^{t}H_{e}^{I}(\tau )d\tau
+\int_{0}^{t}H_{e}^{I}(t_{1})\int_{0}^{t_{1}}H_{e}^{I}(t_{2})dt_{2}dt_{1}+O(%
\delta ^{2})$ is equivalent to the evolution in interaction picture, which
could characterize the noisy evolution. The generator of $U_{e}$ is $%
H_{e}^{I}=U_{0}^{\dagger }H_{e}U_{0}$, where $H_{e}=-\frac{\delta }{2}\sigma
_{z}$.$\ \delta $ being small means $g\ll \Delta $ as derived in supplement,
then $\frac{2g}{\Upsilon _{+}},\frac{-\Delta +\sqrt{4g^{2}+\Delta ^{2}}}{%
\Upsilon _{-}}\rightarrow 0$, $\frac{\Delta +\sqrt{4g^{2}+\Delta ^{2}}}{%
\Upsilon _{+}},\frac{2g}{\Upsilon _{-}}\rightarrow 1$, hence the dipole
moments could be considered homogeneous. As a conclusion, correcting errors
in this regime only concerns about the error $\delta $ in $\sigma _{z}$. 
\begin{figure}[t]
\centering\includegraphics[width=1\columnwidth]{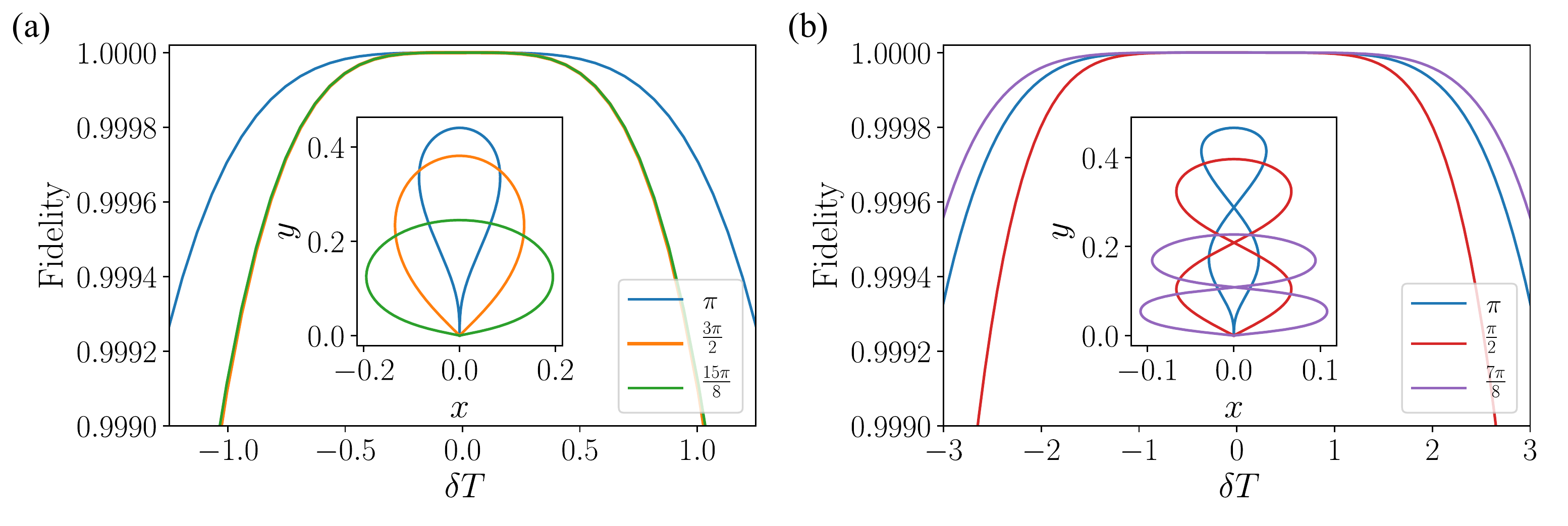}
\caption{Trajectories in parametrization space fidelities and of robust
gates $\{\protect\pi ,\frac{\protect\pi }{2},\frac{\protect\pi }{8}\}$
corresponding to colors blue, orange and green in (a) and (b). (a) shows the
results of first order correction gates while (b) shows the results for
correction gates up to second order errors. The high fidelity plateaus are
generally above $0.99999$ over a variation range of $\protect\delta T=1$ for
1$^{st}$ order cancellation and $\protect\delta T=2$ for 2$^{nd}$ order
cancellation, with gate time $50$ ns simulated for quantum dots. }
\label{Fig_1Qrobust}
\end{figure}

Since the operation $\frac{H_{e}^{I}(t)}{-\delta/2}=U_{0}^{\dagger}\sigma
_{z}U_{0}$ is unitary with matrix norm $\left\vert \left\vert \frac{%
H_{e}^{I}(t)}{-\delta/2}\right\vert \right\vert =1$, it could be mapped onto
a point on a Bloch-like sphere \cite{barnes2015robust}. The difference here
is that this sphere represents the geometry of an operator instead of qubit
states. So we parametrize $\frac{H_{e}^{I}(t)}{-\delta/2}=\overrightarrow{n}%
(t)\cdot\overrightarrow{\sigma}$, where $\overrightarrow{n}(t)$ is an analog
of Bloch vector. The change of $\frac{H_{e}^{I}(t)}{-\delta/2}$ in terms of
time could be characterized as a directional trajectory $\overrightarrow{r}%
(t_{0}:t)$ on the parametrization sphere, starting from $\overrightarrow{r}%
(t_{0})$ and ending at $\overrightarrow{r}(T)$ as finishing a gate. So $%
\left\vert \overset{\cdot}{\overrightarrow{r}}(t)\right\vert =\left\vert
\left\vert \frac{H_{e}^{I}(t)}{-\delta/2}\right\vert \right\vert =1$, which
defines time $t$ in terms of the spherical parameters 
\begin{equation}
t=\int_{_{0}}^{\overrightarrow{r}}d\overrightarrow{r}^{\prime}.  \label{Eq_t}
\end{equation}
Also, since $\overset{\cdot\cdot}{\overrightarrow{r}}(t)$ $\cdot 
\overrightarrow{\sigma}=\frac{i}{\delta/2}U_{0}^{\dagger}[H_{0},H_{e}]U_{0}$%
, the curvature of the curve $\overrightarrow{r}(t_{0}:t)$ is proven \cite%
{supp} to define the control field 
\begin{equation}
\Omega(t)=\overset{\cdot\cdot}{\overrightarrow{r}}(t).  \label{Eq_Omega}
\end{equation}
Further, the torque of $\overrightarrow{r}$ \cite{lehto2015geometry} 
\begin{equation}
\overset{\cdot}{\alpha}(t)=\frac{(\overset{\cdot}{\overrightarrow{r}}\times%
\overset{\cdot\cdot}{\overrightarrow{r}})\cdot\overset{\cdot\cdot \cdot}{%
\overrightarrow{r}}}{\left\vert \left\vert \overset{\cdot }{\overrightarrow{r%
}}\times\overset{\cdot\cdot}{\overrightarrow{r}}\right\vert \right\vert ^{2}}
\label{Eq_theta}
\end{equation}
defines the rotational angle for the gate operation. Therefore, the overall
evolution trajectory $U_{0}(t)$ generated by time-dependent Hamiltonian $%
H_{0}(t)$ could be fully determined in terms of Eq.\ref{Eq_t}to\ref{Eq_theta}%
. Furthermore the pulse constraints for the ERG pulses could be determined.
Since the gate fidelity subjects to errors $F(U_{0}U_{e},U_{0})=\frac{1}{2}%
\left\vert Tr(U_{e})\right\vert $, a ERG requires at $t=T$, $\frac {%
\partial(\left\vert Tr(U_{e})\right\vert )}{\partial\delta}\rightarrow0$.
The first order correction gives $\partial(\left\vert \frac{\delta}{2}%
Tr(\int_{0}^{T}dt\overset{\cdot}{\overrightarrow{r}}(t)\cdot\overrightarrow{%
\sigma })\right\vert )/\partial\delta=0$. Similarly, the second order
correction requires $\int_{0}^{T}dt_{1}\overrightarrow{r}(t_{1})\times%
\overset{\cdot }{\overrightarrow{r}}(t_{1})=0$, so on and so forth.
Therefore, all the information about the parameter $\overset{}{%
\overrightarrow{r}}(t)$ paves the way to look for any ERG pulses up to
arbitrary order of correction. Our related work ref.\cite{YJHai2021construct}
presents systematic construction of the ERG pulses.

\begin{figure}[tb]
\centering\includegraphics[width=1\columnwidth]{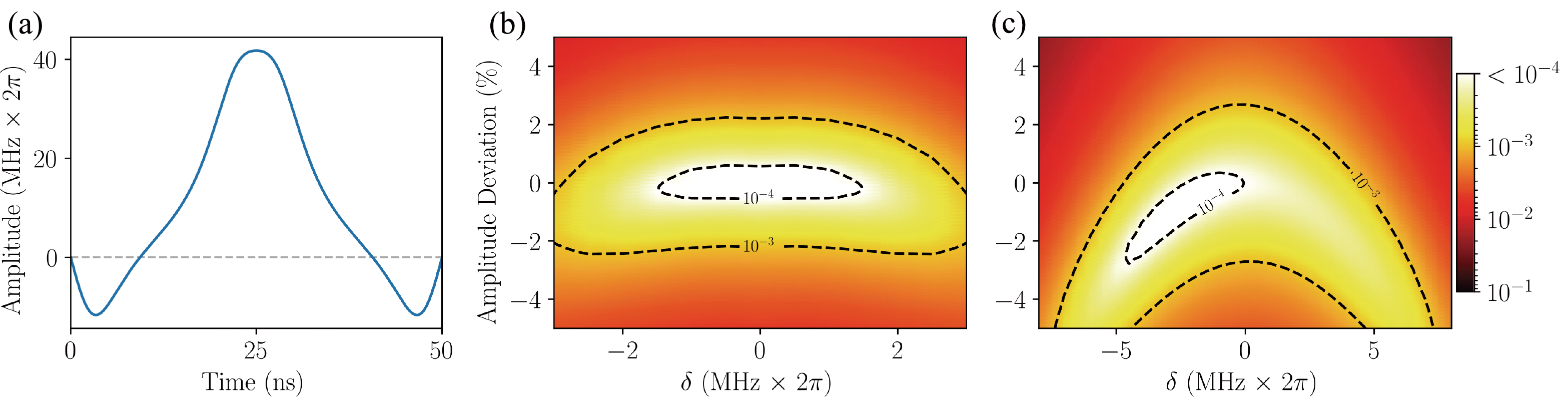}
\caption{Application of a first order $\protect\pi $ RG pulse, see (a), is
applied to superconducting transmon qubit system to implement $X$ gate and
iSWAP gate, with the numerical simulation results shown in (b) and (c),
respectively. The first order RG pulse is resolved from the $\protect\pi $
parametrized curve in Fig.\protect\ref{Fig_1Qrobust} (a). The parameters are 
$\protect\omega _{s}=5$ GHz$\times 2\protect\pi $, $\protect\alpha _{s}=230$
MHz$\times 2\protect\pi $, $\protect\omega _{t}=5.5$ GHz$\times 2\protect\pi 
$, $\protect\alpha _{t}=260$ MHz$\times 2\protect\pi $. The $XX+YY$
interaction strength $g$ varies in the range $-58<g<58$ MHz$\times 2\protect%
\pi $ and $-100<g<100$ MHz$\times 2\protect\pi $ to obtains the range of $%
\protect\delta $ in (b) and (c). Four levels are simulated for the
transmons. }
\label{Fig_robustSCQ}
\end{figure}

As simple examples demonstrating the performance of\ ERG, we set $\alpha=0$.
Now $\overrightarrow{r}(t_{0}:t)$ reduces to a directional curve on a two
dimensional plane \cite{zeng2018general}. And the ideal rotational angle $%
\phi$ is reduced to this constraint%
\begin{equation}
\phi=\arctan\left\vert \frac{\overset{\cdot}{\overrightarrow{r}}(T)\times%
\overset{\cdot}{\overrightarrow{r}}(0)}{\overset{\cdot }{\overrightarrow{r}}%
(T)\cdot\overset{\cdot}{\overrightarrow{r}}(0)}\right\vert =\Delta\varphi+\pi%
\text{ or }\Delta\varphi
\end{equation}
We then find a universal gate set of ERG to implement single qubit X
rotation with angles $\pi$, $\pi/2$ and $\pi/8$, see Fig.\ref{Fig_1Qrobust}.
The fidelity is numerically calculated in a TLS with a range of fluctuation $%
\delta$ in $\sigma_{z}$. Wide range of high fidelity plateaus ($F>0.99999$)
are demonstrated both for first order and second order ERG. These plateaus
are never obtained in commonly used pulses, such as Gaussian, Cosine, etc.
Here, Cartesian coordinates $x=r\cos\theta$, $y=r\sin\theta$ are used to
plot the parametrized curve, see the inset in Fig.\ref{Fig_1Qrobust}(a) and
(b). Then we apply the $\pi$ rotation curve to generate X gate as well as
two qubit iSWAP gate for superconducting transmon qubits. Based on
experimental parameters, the numerical results shown in Fig.\ref%
{Fig_robustSCQ} also illustrate high fidelity plateaus for both single qubit
operation and iSWAP. The ERG also exhibits certain robustness over pulse
amplitude deviation ($1$ to $2\%$ as shown in Fig.\ref{Fig_robustSCQ}(b) and
(c)). The iSWAP fidelity plateau shifts to negative $\delta$ because of the
leakage to transmon's higher levels. Note that with the geometric formalism,
finding appropriate pulses is still a challenging task and requires some
techniques to systematical construction \cite{YJHai2021construct}.

\textit{Discussion.-}We have reported very promising approaches to correct
spatial correlated errors in multi-qubit systems. We have firstly analyzed
the mechanism of gate errors in multi-qubit system subject to spatial
correlation. We have characterized two major regimes of couplings, and
further studied in details the errors associated with spectators and
intruders. To correct the errors induced by intruders, we propose
targeted-correction gates to overcome not only level-splitting issue but
also inhomogeneous control amplitudes. These two issues are by-passed by
previous works. However, in some quantum computation or simulation tasks,
operation qubit system in this regime provides lots of benefits. Now given
our solutions, people could resolve these two issues potentially and go into
a new paradigm of quantum computing with large always-on interactions. To
correct the errors due to spectators, we propose error robust gates to
produce high fidelity plateaus for a range of parameter variation. Compared
to any conventionally-used pulses, our results lead great advantages on gate
robustness for parameter variation. Also, the smoothness and
limited-bandwidth of our pulses guarantee the waveform distortion under a
controllable limit. For both scenarios, we have demonstrated exciting
results with smooth and analytical pulses, high fidelities and powerful
robustness. These results are supported with numerical simulations based on
experimental parameters, which shows promising application in realistic
multi-qubit quantum processors with current architecture or even allowing
hardware simplifications.

Finally, we summarize some further observations of our approaches: 1. TCG
provides an new paradigm for quantum computing or quantum simulation with
always-on interaction, which is friendly for controlling fixed frequency
qubits without increasing the complexity of control circuits; 2. Using ERG
control, the residual coupling issue on current architecture doesn't need to
be eliminated; 3. TCG and ERG together opens up a new degree of control
freedom; 4. Our approaches reduce the need for precise control of qubit
frequency or tunable coupler. 5. Both TCG and ERG's results have small
number of parameters for further optimization, which is very friendly for
experiments.

This work was supported by the Key-Area Research and Development Program of
Guangdong Province (Grant No. 2018B030326001), the National Natural Science
Foundation of China (Grant No. U1801661), the Guangdong Provincial Key
Laboratory (Grant No. 2019B121203002), the Guangdong Innovative and
Entrepreneurial Research Team Program (Grant No. 2016ZT06D348), the Natural
Science Foundation of Guangdong Province (Grant No. 2017B030308003), and the
Science, Technology, and Innovation Commission of Shenzhen Municipality
(Grants No. JCYJ20170412152620376 and No. KYTDPT20181011104202253).

\bibliographystyle{apsrev4-1}
 \newcommand{\noop}[1]{}
\end{document}